\documentclass[pra,showpacs,twocolumn,floatfix]{revtex4}
\usepackage{graphicx}

\begin{document}
\def\bea{\begin{eqnarray}}
\def\eea{\end{eqnarray}}
\def\ben{\begin{equation}}
\def\een{\end{equation}}
\def\benu{\begin{enumerate}}
\def\enu{\end{enumerate}}

\def\n{n}

\def\sss{\scriptscriptstyle\rm}

\def\g{_\gamma}

\def\l{^\lambda}
\def\lfc{^{\lambda=1}}
\def\lo{^{\lambda=0}}

\def\marnote#1{\marginpar{\tiny #1}}
\def\rsav{\langle r_s \rangle}
\def\invdif{\frac{1}{|\br_1 - \br_2|}}

\def\hatT{{\hat T}}
\def\hatV{{\hat V}}
\def\hatH{{\hat H}}
\def\1var{(\bx_1...\bx\N)}

\def\half{\frac{1}{2}}
\def\quart{\frac{1}{4}}

\def\bp{{\bf p}}
\def\br{{\bf r}}
\def\bR{{\bf R}}
\def\bu{{\bf u}}
\def\bmu{{\bf \mu}}
\def\bx{{\br t}}
\def\by{{y}}
\def\ba{{\bf a}}
\def\bq{{\bf q}}
\def\bj{{\bf j}}
\def\bX{{\bf X}}
\def\bF{{\bf F}}
\def\bchi{{\bf \chi}}
\def\bof{{\bf f}}

\def\cA{{\cal A}}
\def\cB{{\cal B}}

\def\x{_{\sss X}}
\def\c{_{\sss C}}
\def\s{_{\sss S}}
\def\xc{_{\sss XC}}
\def\Hx{_{\sss HX}}
\def\Hc{_{\sss Hc}}
\def\Hxc{_{\sss HXC}}
\def\xj{_{{\sss X},j}}
\def\xcj{_{{\sss XC},j}}
\def\N{_{\sss N}}
\def\H{_{\sss H}}

\def\ext{_{\rm ext}}
\def\pot{^{\rm pot}}
\def\hyb{^{\rm hyb}}
\def\HF{^{\rm HF}}
\def\hah{^{1/2\& 1/2}}
\def\loc{^{\rm loc}}
\def\LSD{^{\rm LSD}}
\def\LDA{^{\rm LDA}}
\def\GEA{^{\rm GEA}}
\def\GGA{^{\rm GGA}}
\def\SPL{^{\rm SPL}}
\def\sce{^{\rm SCE}}
\def\PBE{^{\rm PBE}}
\def\DFA{^{\rm DFA}}
\def\TF{^{\rm TF}}
\def\VW{^{\rm VW}}
\def\helm{^{\rm unamb}}
\def\una{^{\rm unamb}}
\def\ion{^{\rm ion}}
\def\HOMO{^{\rm HOMO}}
\def\gs{^{\rm gs}}
\def\dyn{^{\rm dyn}}
\def\adia{^{\rm adia}}
\def\I{^{\rm I}}
\def\pot{^{\rm pot}}
\def\sav{^{\rm sph. av.}}
\def\syv{^{\rm sys. av.}}
\def\pnav{^{\rm sym}}
\def\av#1{\langle #1 \rangle}
\def\unif{^{\rm unif}}
\def\LSD{^{\rm LSD}}
\def\ee{_{\rm ee}}
\def\vir{^{\rm vir}}
\def\ALDA{^{\rm ALDA}}
\def\VUC{^{\rm VUC}}
\def\PGG{^{\rm PGG}}
\def\GK{^{\rm GK}}
\def\atom{^{\rm atmiz}}
\def\trans{^{\rm trans}}
\def\SPA{^{\rm SPA}}
\def\SMA{^{\rm SMA}}

\def\sav{^{\rm sph. av.}}
\def\syv{^{\rm sys. av.}}

\def\up{_\alpha}
\def\dn{_\beta}
\def\up{_\uparrow}
\def\dn{_\downarrow}

\def\td{time-dependent~}
\def\KS{Kohn-Sham~}
\def\DFT{density functional theory~}

\def\fourint{ \int_{t_0}^{t_1} \! dt \int \! d^3r\ }
\def\fourintp{ \int_{t_0}^{t_1} \! dt' \int \! d^3r'\ }
\def\intx{\int\!d^4x}
\def\sph_int{ {\int d^3 r}}
\def\radint{ \int_0^\infty dr\ 4\pi r^2\ }

\def\PRA{Phys. Rev. A\ }
\def\PRB{Phys. Rev. B\ }
\def\PRL{Phys. Rev. Letts.\ }
\def\JCP{J. Chem. Phys.\ }
\def\JPCA{J. Phys. Chem. A\ }
\def\IJQC{Int. J. Quant. Chem.\ }

\def\z{_\zeta}
\def\semi{^{\rm semi}}
\def\slw{_{\rm semi}}
\def\F{_{\sss F}}
\def\loc{^{(0)}}
\def\cE{{\cal E}}
\def\QC{^{\rm QC}}

\title{Exact condition on the Kohn-Sham kinetic energy, and modern parametrization of the Thomas-Fermi density}
\author{Donghyung Lee}
\author{Kieron Burke}
\affiliation{Department of Chemistry, University of California, Irvine, California  92697, USA}
\author{Lucian A. Constantin}
\author{John P. Perdew}
\affiliation{Department of Physics and Quantum Theory Group, Tulane University, New Orleans, 
Louisiana 70118, USA}
\date{\today}
\begin{abstract}
We study the asymptotic expansion of the neutral-atom energy
as the atomic number $Z \to \infty$,
presenting a new method to extract the coefficients from oscillating numerical data.
We find
that recovery of the correct expansion is an exact condition on the Kohn-Sham kinetic energy that is important 
for the accuracy of approximate kinetic energy
functionals for atoms, molecules and solids, when evaluated on a Kohn-Sham density.
For example, this determines the 
small gradient limit of any generalized gradient approximation, and conflicts somewhat with the
standard gradient expansion.
Tests are performed on atoms, molecules, and jellium clusters.
We also give a modern, highly accurate
parametrization of the Thomas-Fermi density of neutral atoms.
\end{abstract}


\maketitle

\section{Introduction}
\label{s:intro}
Ground-state Kohn-Sham (KS) density functional theory (DFT) 
is a widely-used tool for electronic structure
calculations of
atoms, molecules, and solids \cite{FNM03}, in which
only the density
functional for the exchange-correlation energy,
$E\xc[\n]$, must be approximated.
But a direct, orbital-free density functional theory
could be constructed if only the non-interacting
kinetic energy, $T\s$, were known sufficiently accurately
as an explicit functional of the density \cite{DG90}.  Using it would lead
automatically to an electronic structure method
that scales linearly with the number of electrons $N$ 
(with the possible exception
of the evaluation of the Hartree energy). 
Thus the KS kinetic energy functional is something of a holy
grail of density functional purists, and interest
in it has recently revived \cite{KJTH08}.

In this work, we exploit the ``unreasonable accuracy" of 
asymptotic expansions \cite{Sa80}, in this case for
large neutral atoms,  to show that there is a very simple exact
condition that approximations to $T\s$ must satisfy, if
they are to attain high accuracy for total energies of
matter.  By matter, we mean all atoms, molecules,
and solids that consist of electrons in the field of
nuclei, attracted by a Coulomb potential.  
The exact condition is the (known) asymptotic expansion of
$T\s/Z^{7/3}$ for neutral atoms, in powers of $Z^{-1/3}$.
By careful extrapolation from accurate numerical calculations
up to $Z\sim 90$, we calculate the coefficients of this expansion. We find
that the usual gradient expansion,
derived from the slowly-varying gas, but applied to essentially exact
densities, yields only a good approximation to these coefficients.
Thus, all new approximations should either build in these 
coefficients, or be tested to see how well they approximate them.
We perform several tests, using atoms, molecules, jellium surfaces,
and jellium spheres, and analyze two existing approximations.
In Ref. \cite{PCSB06}, a related method was used to derive
the gradient coefficient in modern generalized gradient approximations (GGA's)
for exchange. 
Given this importance of $N=Z \to \infty$ as a condition on functionals,
we revisited and improved upon the existing parametrizations of the 
neutral-atom Thomas-Fermi (TF)
density. The second-half of the paper is devoted to testing its accuracy.
\section{Theory and Illustration}
\label{s:theory}

For an $N$-electron system, the Hamiltonian is 
\ben
\hat{H} = \hat{T} + \hat{V}\ext + \hat{V}\ee\,,
\label{eq:Hamil}
\een
where $\hat{T}$ is the kinetic energy 
operator, $\hat{V}\ext$ the external potential, and $\hat{V}\ee$ the electron-electron 
interaction, respectively. The electron density $n(\br)$ yields 
$N=\int d^{3}r\, n(\br)$, where $N$ is the particle number.

To explain asymptotic exactness, we (re)-introduce the $\zeta$-scaled potential \cite{L81} 
(which is further discussed in Ref. \cite{ELCB08}),
given by 
\ben
v\ext^\zeta(\br) = \zeta^{4/3}\, v\ext(\zeta^{1/3} \br),~~~~~~N\to\zeta N,
\label{vzeta}
\een
where $v\ext(\br)$ is the external potential, and the Thomas-Fermi expectation
value is $V\ext^{\zeta}[n]=\zeta^{7/3}V\ext[n]$.
In this $\zeta$-scaling scheme, nuclear positions ${\bf R}_\alpha$
and charges $Z_\alpha$ of molecules are scaled into 
$\zeta^{-1/3} {\bf R}_\alpha$ and $\zeta Z_\alpha$ respectively.  
In a uniform electric
field, ${\mathcal E} \to  \zeta^{5/3} {\mathcal E}$. For neutral atoms, 
scaling $\zeta$ is the same as scaling $Z$, and this gives Schwinger's 
asymptotic expansion for the total energy of  neutral atoms \cite{Sa80, ES85},
\ben
E = - c_0\, Z^{7/3} - c_1\, Z^2 - c_2\, Z^{5/3} + \cdots\,,
\label{EZasymp}
\een
where $c_0=0.768745$, $c_1=-1/2$, $c_2=0.269900$, and $Z$ is the atomic number.
This large $Z$-expansion gives a remarkably good approximation to the 
Hartree-Fock energy of the neutral atoms, with less than a 10\% error for H, and less 
than 0.5\% error for Ne.
By the virial theorem for neutral atoms, $T=-E$, and $T \simeq T\s$ to this order 
in the expansion (since the correlation energy is roughly $\sim Z$).
Hence, the non-interacting kinetic energy has the following asymptotic expansion.
\ben
T\s = c_0\, Z^{7/3} + c_1\, Z^2 + c_2\, Z^{5/3} + \cdots
\label{TsZasymp}
\een

We say that an approximation to the kinetic energy functional is 
{\em asymptotically exact} to the $p$-th degree 
if it can reproduce the exact $c_{0}, c_{1},\ldots,c_{p}$.  
The three displayed terms in Eq. (\ref{EZasymp}) constitute the second-order asymptotic expansion
for the total energy of neutral atoms, and we expect that this asymptotic 
expansion is a better starting point for constructing a more accurate 
approximation to the kinetic
energy functional than the traditional gradient expansion approximation (GEA). 

\begin{figure}[htbp]
\begin{center}
\includegraphics[height=5.5cm]{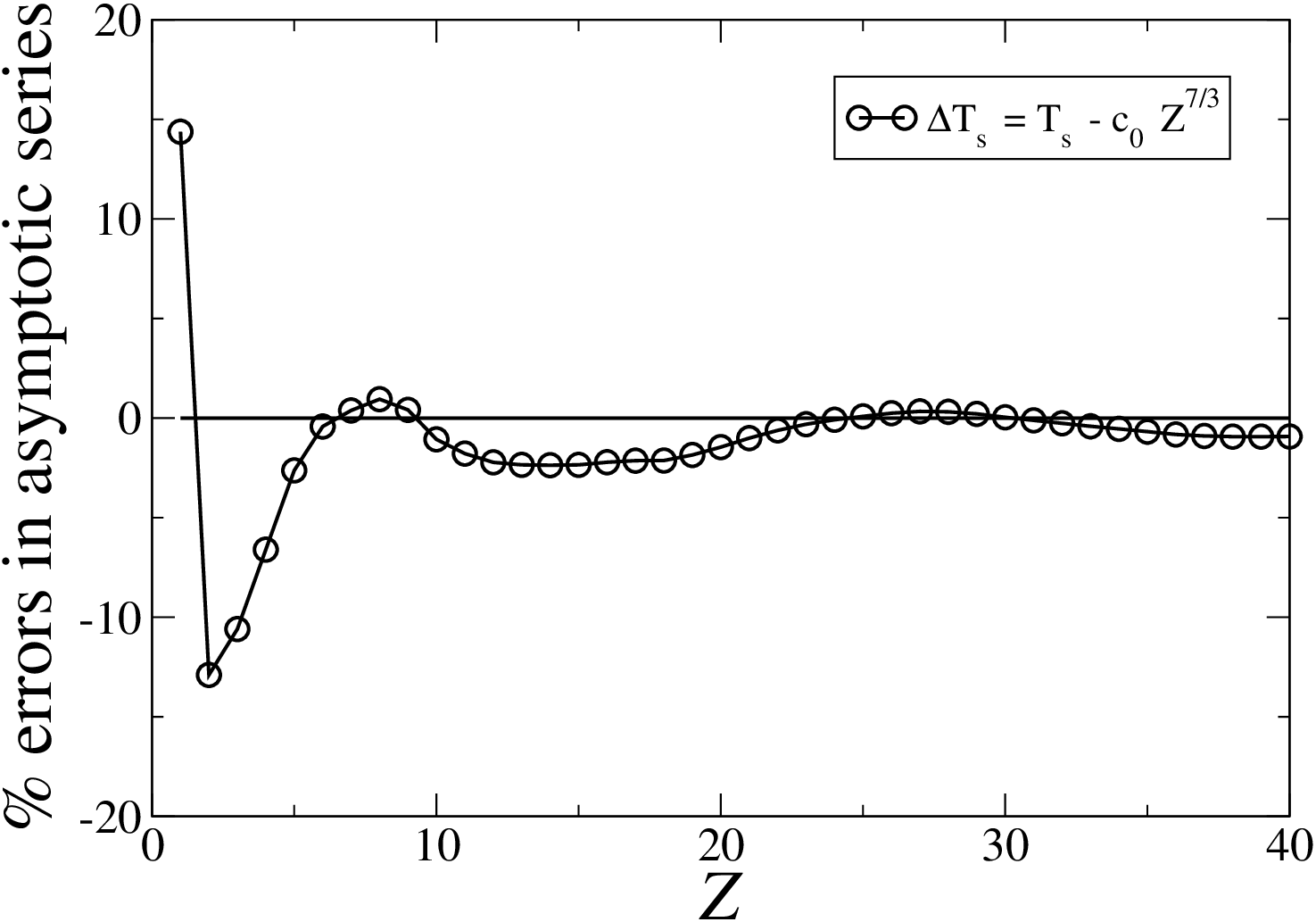}
\caption{Percentage error between $c_1\, Z^2 + c_2\, Z^{5/3}$ and $\Delta T\s = T\s-c_{0}Z^{7/3}$.}
\label{f:deltaTs}
\end{center}
\end{figure}

The leading term in Eq. (\ref{TsZasymp}) is given \emph{exactly} by a local approximation to $T\s$ (TF theory), 
but the leading \emph{correction} is due to higher-order quantum effects, and
only approximately given 
by the gradient expansion evaluated on the \underline{exact} density.
However, these coefficients are \emph{vital} to finding accurate kinetic energies.
Since we know that $c_{0}Z^{7/3}$ becomes exact 
as $N=Z \to \infty$, we define $\Delta T\s = T\s - c_{0}Z^{7/3}$ 
and investigate $\Delta T\s$ as a function of $Z$.
How accurate is the asymptotic expansion for $\Delta T\s$?
In Figure \ref{f:deltaTs}, we evaluate $T\s$ for atoms
within the optimized effective potential (OEP) \cite{TS76} using the exact exchange 
functional and plot the percentage error in $\Delta T\s$,
for all atoms and the asymptotic 
series with just two terms. The series is incredibly accurate,
with only a 13\% error for $N$=2 (He), and 14\% for $N$=1.
Thus, any approximation that reproduces the correct asymptotic series
(up to and including the $c_{2}$ term) is likely to produce a highly accurate $T\s$.
\vspace{0.6cm}

To demonstrate the power and the significance of this approach, we apply it directly to
the first term (where the answer is already known,
but perhaps not yet fully appreciated
in the DFT community).  Using any (all-electron) electronic structure code,
one calculates the total energies of atoms for a sequence running down
a column.  By sticking with a specific column, one reduces the oscillatory
contributions across rows, and the alkali-earth column
yields the most accurate results.   
By then fitting the resulting
curve of $T\s/Z^{7/3}$ as a function of $Z^{-1/3}$ to a parabola,
one finds $c_0=0.7705$.  
Now assume one wishes to make 
a local density approximation (LDA) to $T\s$, but knows nothing about the uniform
electron gas.  Dimensional analysis yields \cite{LP85}
\ben
T\loc[\n] = A\s\, I,~~~I= \int d^3r\, n^{5/3}(\br)\, ,
\label{eq:Tloc}
\een
but does not determine the constant, $A\s$.  A similar fitting of
$I$, based on the self-consistent densities evaluated using the OEP exact exchange 
functional, gives a leading term of 0.2677 $Z^{7/3}$, yielding
$A\s = 2.868$. Thus we have deduced the local approximation
to the non-interacting kinetic energy.

A careful inspection of the above argument reveals that  the
uniform electron gas is never mentioned.
As $N$ grows, the wavelength of the majority of the
particles becomes short relative to the scale on which the potential
is changing, loosely speaking, and semiclassical behavior dominates.  
The local approximation is
a universal semiclassical result, which is exact for a uniform gas
simply because that system has a constant potential.
On the basis of that argument,
we know the exact value is 
$A\s$=$(3/10)(3\pi^{2})^{2/3}=2.871$, demonstrating that (for this
case) our result is accurate to about 0.1\%.
This argument tells us that the reliability of the local approximation 
is no indicator of how rapidly the density varies.
That this argument is correct for neutral atoms was carefully proven
by Lieb and Simon in 1973 \cite{LS73} and later generalized by Lieb to all matter \cite{L81}. 

The focus of the first part of this paper is on the remaining two known coefficients ($c_{1}$ and $c_{2}$)  
and how well the GEA performs for them. 
We evaluate those gradient terms by fitting
asymptotic series exactly and we find that the gradient expansion does well, but is not exact. 
From this information, we develop a modified gradient expansion approximation that
reproduces the exact asymptotic coefficients $c_{1}$ and $c_{2}$,
merely as an illustration of the power of asymptotic exactness.
We test it on a variety of systems, finding the expected behavior.

In Section \ref{sec:MTF}, we present
a parametrization of the TF density which is more 
accurate than previous parametrizations.
The TF density has a simple scaling with $Z$ and 
becomes relatively exact and slowly-varying for a neutral atom 
as $Z \rightarrow \infty$, breaking down only near the nucleus 
and in the tail. We compare various quantities of our parametrization 
with exact values and earlier parametrizations, and analyze the properties of the TF density.

\section{Large $Z$ Methodology}
\label{sec:meth}

We begin with a 
careful methodology for extracting the asymptotic behavior from highly
accurate numerical calculations.  
Fully numerical DFT calculations were performed using the OPMKS 
code \cite{OPMKS} to calculate the total energies of neutral 
atoms using the OEP exact exchange functional.
The spin-density functional version of $T\s$ has been used for all systems \cite{OP79}.
\vspace{0.6cm}
\begin{figure}[htbp]
\begin{center}
\includegraphics[height=5.5cm]{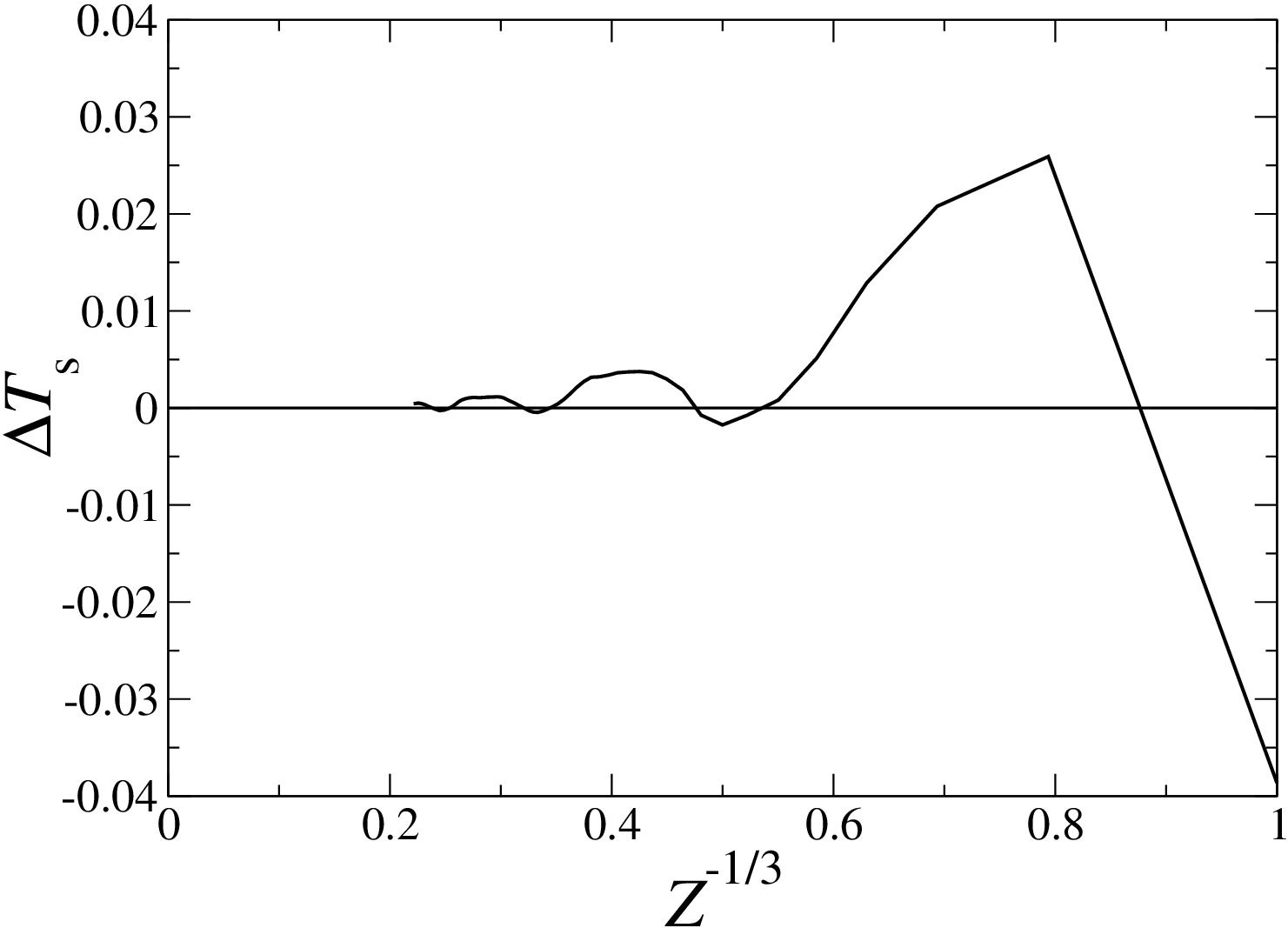}
\caption{Difference between $T\s/Z^{7/3}$ and $c_{0} + c_{1}\, Z^{-1/3} + c_{2}\, Z^{-2/3}$ as a function of $Z^{-1/3}$ with exact asymptotic coefficients.}
\label{f:deltaOsc}
\end{center}
\end{figure}

To attain maximum accuracy for $c_{1}$ and $c_{2}$,
we need to suppress the oscillations
which come from the next term, $c_{3}Z^{4/3}$. 
Consider first the OEP results.
We investigate the differences between $T\s^{\rm OEP}/Z^{7/3}$ and
$c_{0} + c_{1}\, Z^{-1/3} + c_{2}\, Z^{-2/3}$ with exact asymptotic coefficients
in Figure \ref{f:deltaOsc}.
We extract 6 data points ($Z$=24 (Cr), 25 (Mn), 30 (Zn), 31 (Ga), 61 (Pm), and 74 (W))
which have the smallest
differences, i.e., nearest to where the curve crosses the horizontal axis. We then make a least-squares
fit with a parabolic form in $Z^{-1/3}$, ignoring the oscillation term,
\ben
\frac{T}{Z^{7/3}} = 0.768745+c_{1}Z^{-1/3}+c_{2}Z^{-2/3}\, .
\label{eq:fitting}
\een
Effectively, we solve two linear equations for $c_{1}$ and $c_{2}$.
We explicitly include the exact $c_{0} = 0.768745$, since
we don't have enough data points to extract $c_{0}$ accurately, especially in the region $Z^{-1/3} < 0.2$. 
It is important to control the behavior of the fitting line at $Z \to \infty$.
This fitting yields a good estimate of $c_1 = -0.5000$ and $c_{2} = 0.2702$, with error
less than 1\%.  This demonstrates the accuracy of our method for
$c_1$ and $c_{2}$ (by construction).

We repeat the same procedure to 
extract $c_1$ and $c_2$ coefficients of TF and second- and fourth-order GEA's 
which are given by
\ben
T^{\rm GEA2}= T\TF + T^{(2)},
\een
and \cite{Kc57, DG90, H73}:
\ben
T^{\rm GEA4}= T\TF + T^{(2)}+T^{(4)}\,.
\een
These gradient corrections to the local approximation are given by
\ben
T^{(2)} = \frac{5}{27} \int d^3r\, \tau\TF(\br) s^{2}(\br)\,, 
\label{T2}
\een
and
\bea
T^{(4)} =\frac{8}{81} \int d^{3}r\, \tau\TF(\br) \left[ q^{2}(\br)
-\frac{9}{8}q(\br)s^{2}(\br)+\frac{s^{4}(\br)}{3}\right],
\label{T4}
\eea
where $\tau\TF(\br)$, $s(\br)$, and $q(\br)$ are defined as
\ben
\tau\TF(\br)=\frac{3}{10} k\F^{2}(\br)n(\br)\,,
\label{t0}
\een
\ben
s(\br)= \frac{|\nabla n(\br)|}{2k\F(\br) n(\br)} \,,
\label{s}
\een
\ben
q(\br)=\frac{\nabla^{2} n(\br)}{4k\F^{2}(\br) n(\br)}\,,
\label{q2}
\een
and $ k\F(\br) = (3\pi^{2}n(\br))^{1/3}$.

We have also applied this procedure
to both $T^{(2)}$ and $T^{(4)}$.
Since the asymptotic expansions of these energies begin at $Z^2$, we 
extract only a $c_1$ and a $c_2$ for each using the following equations:
\bea
\frac{T^{\rm GEA2} - T\TF}{Z^{7/3}} &=& \Delta c_{1}Z^{-1/3}+ \Delta c_{2}Z^{-2/3}\, , \nonumber \\
\frac{T^{\rm GEA4} - T^{\rm GEA2}}{Z^{7/3}} &=& \Delta c_{1}Z^{-1/3}+ \Delta c_{2}Z^{-2/3}\, .
\eea
These results are also included in Table \ref{t:CoeffAlea}, and are
of course consistent with our results from Eq. (\ref{eq:fitting}).
\begin{table}[htb]
\begin{tabular}{|c|c|c|}
\hline
&  $c_1$  & $c_2$\\
\hline
  Exact&   -0.5000&     0.2699\\
  $T^{\rm OEP}$&    -0.5000&  0.2702\\
  $T\TF$&   -0.6608&   0.3854\\
  $T^{\rm (2)}$&   0.1246  & -0.0494 \\
  $T^{\rm (4)}$&  0.0162& 0.0071 \\
  $T^{\rm GEA2}$&  -0.5362& 0.3360\\
  $T^{\rm GEA4}$&  -0.5200& 0.3431\\
  $T^{\rm GGA}$\footnotemark[1]&  -0.5080 & 0.2918\\
  $T^{\rm LmGGA}\footnotemark[1]$&  -0.5089 & 0.3174\\
\hline
\end{tabular}
\footnotetext[1]{See section \ref{s:interp}}
\caption{\label{t:CoeffAlea} The coefficients in the asymptotic expansion of
the exact kinetic energy and various local and semilocal functionals.
The fit was made to $Z$=24 (Cr), 25 (Mn), 30 (Zn), 31 (Ga), 61 (Pm),  and 74 (W).
The functionals of the last two rows are defined in section \ref{s:interp}.}
\end{table}

\section{Results and Interpretation}
\label{s:interp}

To understand the meaning of the above results, begin with
the values of $c_1$.  
We have combined the results of the $T^{(2)}$ and $T^{(4)}$ fits with that of the $T\TF$ fit
to produce the asymptotic coefficients of $T^{\rm GEA2}$ and $T^{\rm GEA4}$.
We check that these combinations produce the same coefficients in Table \ref{t:CoeffAlea} which
are found from the direct fitting of $T^{\rm GEA2}$ and $T^{\rm GEA4}$ using Eq. (\ref{eq:fitting}).
The exact
value of $c_1$ is $-1/2$.  We see that the local
approximation (TF) gives a good estimate, $-0.66$.  Then the second-order
gradient expansion yields
$-0.54$, reducing
the error by a factor of 5.  Finally, the fourth-order gradient expansion
yields $-0.52$, a further
improvement, yielding only a 4\% error in its approximation to the Scott correction \cite{S52}.

For $c_2$, the gradient expansion is
less useful.  The exact result is $0.27$, while the TF
approximation overestimates this as 0.39. The GEA2 result is only
slightly reduced (0.34), and the fourth-order correction has the
wrong sign.

To understand how important these results can be, we consider how
exchange and correlation functionals are constructed.  Often, such
constructions begin from the GEA, which is
then generalized to include (in an approximate way) all powers of a given gradient.
For slowly varying densities, it is considered desirable to
recover the GEA result.   But we have seen here how this conflicts
with the asymptotic expansion, and in Ref \cite{PCSB06}, it was shown how the
asymptotic expansion is more significant to energies of real 
materials, and how successful GGA's for atoms and molecules well-approximate
the large-$Z$ asymptotic
result, not the slowly-varying gas.

{\bf Atoms:} To illustrate this point, we construct here a trivial modified gradient
expansion, MGEA2, designed to have the correct asymptotic coefficients,
in so far as is possible.  Thus
\ben
T^{\rm MGEA2} = T\TF+1.290\, T^{(2)}
\label{eq:MGEA2}
\een
The enhancement coefficient has been chosen to make $c_1^{\rm MGEA2}=-1/2$
exactly.  
In Table \ref{t:TAl-ea}, we list the results of several different approximations
for the alkali-earth atoms.
Because the GEA2 error passes through 0 around Z=8, its errors are
artificially low.
\begin{table*}[htb]
\begin{tabular}{|c|c|c|cc|cc|cc|cc|cc|}
\hline
Atom & Z & $T^{\rm OEP}$ & $T\TF$ & \%err & $T^{\rm GEA2}$ & \%err &   $T^{\rm MGEA2}$ & \%err &$T^{\rm GEA4}$ & \%err & $T^{\rm MGEA4}$ & \%err \\ 
\hline
Be & 4&14.5724 & 13.1290 & -10 & 14.6471 & 0.5 & 15.0880 & 3.5 & 14.9854 & 2.8 &  14.5453 & -0.2\\ 
Mg &12& 199.612 & 184.002 & -8 & 198.735 & -0.4 &  203.014 & 1.7 &201.452 & 0.9 & 199.924 & 0.2\\ 
Ca &20& 676.752 & 630.064 & -7 & 672.740 & -0.6 &  685.136 & 1.2 & 680.286 & 0.5 & 677.433 & 0.1\\ 
Sr &38& 3131.53 & 2951.89 & -6 & 3110.44 & -0.7 &  3156.50 & 0.8 &3136.76 & 0.2 & 3134.48 & 0.09\\ 
Ba &56& 7883.53 & 7478.27 & -5 & 7829.36 & -0.7 &  7931.34 & 0.6 &7886.19 & 0.03 & 7888.14& 0.06\\ 
Ra &88& 23094.3 & 22065.8 & -4 & 22945.9 & -0.6 &  23201.5 & 0.5 &23083.9 & -0.05 & 23110.5 & 0.07\\ 
\hline
\end{tabular}
\caption{\label{t:TAl-ea} KS kinetic energy ($T$)  in hartrees and various approximations for alkali-earth atoms.}
\end{table*}

\begin{figure}[htbp]
\begin{center}
\includegraphics[height=5.5cm]{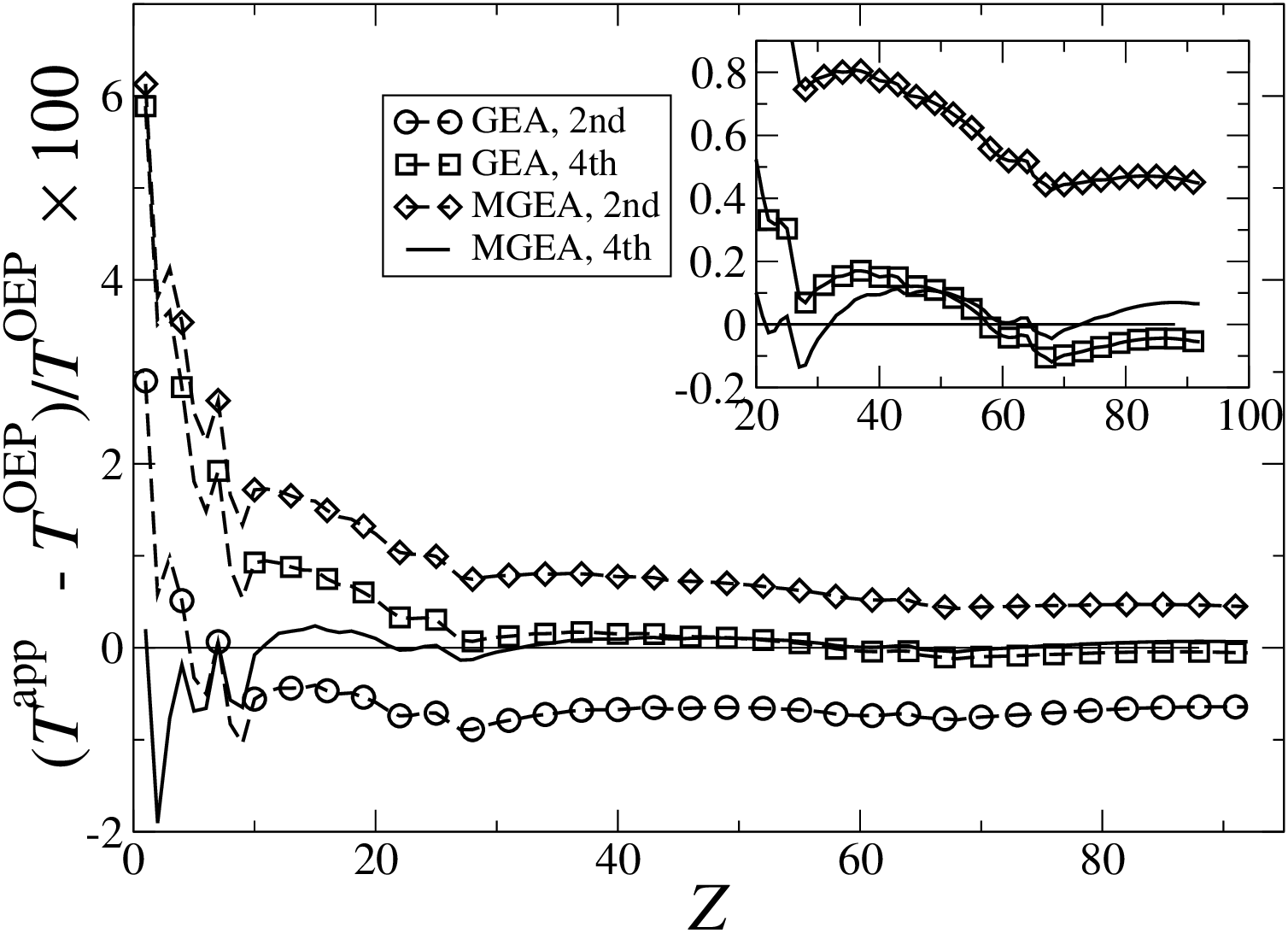}
\caption{Percentage errors for atoms (from $Z=1$ to $Z=92$) using various approximations.}
\label{f:Tallatoms}
\end{center}
\end{figure}

We can repeat this exercise for the fourth order, requiring both
$c_1$ and $c_2$ be exact.  Now we find:
\ben
T^{\rm MGEA4} [\n] = T\TF[\n] + 1.789\, T^{(2)}[\n] - 3.841 T^{(4)}[\n]
\label{eq:MGEA4}
\een
i.e., strongly modified gradient coefficients.  
This is somewhat
arbitrary, as there are several terms in $T^{(4)}$, and there's
no real reason to keep their ratios the same as in GEA (Eq. (\ref{T4})).
However, the results of Table \ref{t:TAl-ea} and Figure \ref{f:Tallatoms} speak for themselves. The
resulting functional is better than either GEA for all the
alkali-earths.
Of course, the exact $T\s$ is positive for any density, as are the terms
$T\TF$, $T^{(2)}$ and $T^{(4)}$ of the GEA.  Eq. (\ref{eq:MGEA4}) however
can be improperly negative for rapidly-varying densities, and so is not suitable for general use.
\vspace{0.6cm}

{\bf Molecules:} The improvement in total kinetic energies is not just confined to atoms.
Also, for non-interacting kinetic energies of molecules, 
using the data in Ref. \cite{IEMS01}, 
Eq. (\ref{eq:MGEA4}) gives 
better average of the absolute errors in hartree (0.6) than $T\TF$ (9.4), $T^{\rm GEA2}$ (0.9), 
and $T^{\rm GEA4}$ (0.8), shown in Table \ref{t:KEAM}. 
\begin{table}[htb]
\begin{tabular}{|c|c|c|c|c|c|c|}
\hline
Atom & Exact\footnotemark[1] & $T\TF$\footnotemark[1] & $T^{GEA2}$\footnotemark[1] & $T^{GEA4}$\footnotemark[1] & $T^{MGEA4}$ \\ 
\hline
H & 0.500 & -0.044 & 0.011 & 0.032 &  -0.026 \\
B & 24.548 & -2.506 & -0.058 & 0.476 & -0.177 \\
C & 37.714 & -3.731 & -0.154 & 0.600 & -0.228 \\
N & 54.428 & -4.993 & -0.097 & 0.904 & -0.078 \\
O & 74.867 & -6.990 & -0.546 & 0.765 & -0.497 \\
F & 99.485 & -9.093 & -0.933 & 0.659 & -0.609 \\
H$_2$ & 1.151 & -0.142 & -0.014 & 0.033 & -0.094 \\
HF & 100.169 & -9.016 & -0.920 & 0.639 & -0.520 \\
H$_2$O & 76.171 & -7.074 & -0.692 & 0.565 & -0.484 \\
CH$_4$ & 40.317 & -3.773 & -0.140 & 0.619 & -0.189 \\
NH$_3$ & 56.326 & -5.292 & -0.400 & 0.587 & -0.331 \\
BF$_3$ & 323.678 & -29.052 & -2.641 & 2.454 & -1.370 \\
CN & 92.573 & -8.940 & -0.687 & 0.978 & -0.570 \\
CO & 112.877 & -10.694 & -0.911 & 1.036 & -0.670 \\
F$_2$ & 199.023 & -18.367 & -2.201 & 0.925 & -1.451 \\
HCN & 92.982 & -8.925  & -0.658 & 1.008 & -0.534 \\
N$_2$ & 109.013 & -10.487 & -0.916 & 0.999 & -0.719 \\
NO & 129.563 & -12.342 & -1.240 & 0.962 & 0.279 \\
O$_2$ & 149.834 & -14.186 & -1.527 & 0.965 & -1.110 \\ 
O$_3$ & 224.697 & -21.636 & -2.699 &1.028 & -2.071 \\
\hline
MAE\footnotemark[2] & & 9.364 & 0.872 & 0.812 & 0.600 \\
\hline
\end{tabular}
\footnotetext[1]{Ref. \cite{IEMS01}}
\footnotetext[2]{Mean absolute error}
\caption{\label{t:KEAM} Exact non-interacting kinetic energy (in hartrees) for molecules,
and errors in approximations.
All values are evaluated on the converged KS orbitals and 
densities obtained with B88-PW91 functionals, 
and the MGEA4 kinetic energies are evaluated using the TF and the GEA data from Ref. \cite{IEMS01}.
}
\end{table}
Of greater importance are energy differences.
For atomization kinetic energies, also using the data in 
Ref. \cite{IEMS01}, $T\TF$ gives the best averaged 
absolute error (0.25), which is worsened by gradient corrections.
Since the GEA does not have the right quantum corrections from the edges, turning points and
Coulomb cores \cite{ELCB08}, GEA does not improve on the atomization process. 
However, the TF kinetic energy functional is always the dominant term.
So, TF gives very good results on the atomization kinetic energies.
But the error (0.29) of Eq. (\ref{eq:MGEA4}) is smaller than that of $T^{\rm GEA2}$ (0.36) and 
$T^{\rm GEA4}$ (0.44).
In either case, Eq. (\ref{eq:MGEA4}) works better for atoms and molecules 
than the fourth-order gradient expansion.
Thus, requiring asymptotic exactness is a useful
and powerful constraint in functional design.

{\bf Jellium surfaces:} We test this MGEA4 functional for jellium surface kinetic energies. 
As shown in Table \ref{t:jellium}, the $T^{(4)}$ term in $T^{\rm GEA4}$ 
improves the jellium surface kinetic energy in comparison to the results 
of $T^{\rm GEA2}$, but Eq. (\ref{eq:MGEA4}) worsens the jellium surface kinetic energies 
due to the strongly modified coefficient of $T^{(4)}$.
This is a confirmation of our general approach. By building in the correct aysmptotic
behavior for atoms, including the Scott correction coming from the $1s$ region,
we \emph{worsen} energetics for systems without this feature.
\begin{table}[htb]
\begin{tabular}{|c|c|c|c|c|c|c|c|}
\hline
$r_{s}$ & Exact & $T\TF$ & $T^{\rm GEA2}$ & $T^{\rm GEA4}$ & $T^{\rm MGEA2}$\footnotemark[1] & $T^{\rm MGEA4}$\footnotemark[2] & $T^{\rm LmGGA}$ \\ 
\hline
2 & -5492.7 &11 & 2.5 & 1.1 & -0.9 & 0.73 & 1.3 \\ 
4 & -139.9 &54 & 22 & 11 & 12 & 36  & 15\\ 
6 &-3.4 & 660& 330  & 180 & 238 & 675  & 280\\ 
\hline
\end{tabular}
\footnotetext[1]{See Eq. (\ref{eq:MGEA2})}
\footnotetext[2]{See Eq. (\ref{eq:MGEA4})}
\caption{\label{t:jellium} Exact jellium surface kinetic energies ($\text erg/\text cm^{2}$)
and \% error, which is $(\sigma^{\rm app}\s-\sigma^{\rm ex}\s)/\sigma^{\rm ex}\s$, of each approximation.}
\end{table}

{\bf Jellium spheres:} We also investigate the kinetic energies of neutral jellium spheres
(with KS densities using LDA
exchange-correlation and with $r\s = 3.9$) from Ref. \cite{PC07}.
The analysis of the results is based upon the
liquid drop model of Refs. \cite{PWE91, ELD94}.
We write
\ben
T\s(r\s,N) = \frac{4}{3}\pi R^{3} \tau\unif(r\s) + 4\pi R^{2}\sigma\s + 2\pi R\gamma\s^{\rm eff}(r\s,N),
\een
where $R$ is the radius of the sphere of uniform positive background.
Since we know
the bulk (uniform) kinetic energy density, $\tau\unif$,
and the surface kinetic energy $\sigma\s$ for a given functional, we can extract
$\gamma\s^{\rm eff}(r\s,N)$ from
this equation, and
\ben
\lim_{N \to \infty} \gamma\s^{\rm eff}(r\s,N) = \gamma\s(r\s)
\een
is the curvature energy of jellium.
We calculate $\gamma\s^{\rm eff}(r\s,N)$ using the TF, GEA, MGEA,
and a Laplacian-level meta-GGA (LmGGA) of Ref. \cite{PC07}, which is
explained further in the following subsection.
From Table \ref{t:gamma}, 
\begin{table}[htb]
\begin{tabular}{|c|c|c|c|c|c|c|}
\hline
 $N$    &  Exact   &  $T^{\rm GEA2}$  &   $T^{\rm GEA4}$ &   $T^{\rm MGEA2}$\footnotemark[1] &  $T^{\rm MGEA4}$\footnotemark[2] &  $T^{\rm LmGGA}$  \\
\hline
  2        &  -1.8  &  1.1   &   2.4   &      1.5     &       -2.8     &       1.9      \\
  8        &  -1.9  &  1.0   &   2.1   &      1.3     &       -2.3     &       -5.1       \\
 18       &  -0.5  &  1.2   &   2.0   &      1.6     &       -0.7     &      -6.4      \\
 58       &  -0.8  &  1.3   &   2.2    &     1.7      &      -1.1      &     -3.2       \\
 92       & -1.7   &  1.2   &   2.0    &    1.5       &     -1.0       &    -1.9        \\
254      & -0.5   &  1.4   &   2.3    &    1.8       &     -0.9       &    -            \\
\hline
\end{tabular}
\footnotetext[1]{See Eq. (\ref{eq:MGEA2})}
\footnotetext[2]{See Eq. (\ref{eq:MGEA4})}
\caption{\label{t:gamma} $10^{4}\times (\gamma\s^{\rm eff}(\br\s,N)-\gamma\s^{\rm TF}(\br\s,N))$ in atomic units vs.
$N=Z$ for neutral jellium spheres with $r\s = 3.93$ with various
functionals.  As $N=Z \to \infty$, $\gamma\s^{\rm eff}$ tends to the curvature kinetic energy
of jellium, $\gamma\s$.}
\end{table}
we observe that: (i) Gradient corrections in GEA worsen $\gamma\s^{\rm eff}$.
(ii) The LmGGA of Ref. \cite{PC07} is even worse than $T^{\rm GEA4}$. (iii) Eq. (\ref{eq:MGEA2})
(which has the right $c_{0}$ and $c_{1}$) is not so good, but better than $T^{\rm GEA4}$.
(iv) Eq. (\ref{eq:MGEA4}) (which has the right
$c_{0}$, $c_{1}$, and $c_{2}$) gives good results.

{\bf Existing approximations:} We suggest that the large-$Z$ 
asymptotic expansion is a necessary condition that an accurate kinetic energy functional 
should satisfy, but is not sufficient. We show this by testing two kinds of semilocal approximations
(GGA and meta-GGA) to the kinetic energy functionals.

Recently, Tran and Wesolowski \cite{TW02} constructed a GGA-type
kinetic energy functional using the \emph{conjointness conjecture}. They found the
enhancement factor by minimizing mean absolute errors of kinetic energies for closed-shell atoms.
We evaluate the kinetic energies of atoms using this
functional ($T^{\rm GGA}$) and extract the asymptotic coefficients shown in Table \ref{t:CoeffAlea}.
This gives a good $c_{1}$ coefficient, with $c_{2}$ close to the exact value, and so is much
more accurate than the GEA's.

Perdew and Constantin \cite{PC07} constructed a LmGGA for 
the positive kinetic energy density $\tau$ that satisfies the local bound 
$\tau \ge \tau_W$, where $\tau_W$ is the von Weizs\"{a}cker kinetic energy density, and tends 
to $\tau_W$ as $r\to0$ in an atom.
It recovers the fourth-order gradient expansion in the slowly-varying limit.
We calculate the asymptotic coefficients 
shown in Table \ref{t:CoeffAlea} for this functional.
These values are better than those of $T^{\rm GEA4}$.
The good $c_{1}$ from $T^{\rm GEA4}$ appears somewhat fortuitous,  since there is
nothing about a slowly-varying density that is relevant to a cusp in the density.
The good Scott correction $c_1$ from the LmGGA comes from correct
physics: LmGGA recovers the von Weizs\"{a}cker kinetic energy density in the $1s$ cusp,
without the spurious but integrable divergences of the integrand of $T^{\rm GEA4}$.

We finish by discussing other columns of the periodic table.  We have also performed all
these calculations on the noble gases. 
In fact, from studies of
the asymptotic series \cite{ESc85}, it is known that the shell-structure occurs
in the next order, $Z^{4/3}$, and that the noble gases are furthest from the asymptotic curves.
But Table \ref{t:TNoble} shows our functionals work almost as well for the
noble gas series.

\begin{table*}[htb]
\begin{tabular}{|c|c|c|cc|cc|cc|cc|cc|}
\hline
Atom &Z& $T^{\rm OEP}$ & $T\TF$ & \%err & $T^{\rm GEA2}$ & \%err &   $T^{\rm MGEA2}$ & \%err &$T^{\rm GEA4}$ & \%err & $T^{\rm MGEA4}$ & \%err \\ 
\hline
He &2& 2.86168 & 2.56051 & -11 & 2.87847 & 0.6 & 2.97083 & 3.8 &2.96236 & 3.5 & 2.80717 & -1.9\\ 
Ne &10& 128.545 & 117.761 & -8 & 127.829 & -0.6 &  130.753 & 1.7 &129.737 & 0.9 & 128.447 & -0.08\\ 
Ar &18& 526.812 & 489.955 & -7 & 524.224 & -0.5 &  534.178 & 1.4 &530.341 & 0.7 & 527.772 & 0.2\\ 
Kr &36& 2752.04 & 2591.20 & -6 & 2733.07 & -0.7 &  2774.27 & 0.8 &2756.72 & 0.2 & 2754.17 & 0.08\\ 
Xe &54& 7232.12 & 6857.94 & -5& 7183.78 & -0.7 &  7278.42 & 0.6 &7236.65 & 0.06 & 7237.85 & 0.08\\ 
Rn &86& 21866.7 & 20885.7 & -4 & 21725.4 & -0.6 & 21969.3 & 0.5 &21857.2 & -0.04 & 21881.7 & 0.07\\ 
\hline
\end{tabular}
\caption{\label{t:TNoble} KS kinetic energy ($T$) in hartrees and various approximations for noble atoms.}
\end{table*}

\section{Modern Parametrization  of Thomas-Fermi Density}
\label{sec:MTF}

Our asymptotic expansion study gives new reasons for studying large $Z$ atoms.
Our approximate functionals were tested on highly accurate densities, but ultimately, self-consistency
is an important and more-demanding test. Any approximate functional yields an
approximate density via the Euler equation.
In this section, we present a new, modern parametrization of the neutral 
atom TF density, which is more accurate than earlier versions \cite{GD79, LB55}.

The TF density of a neutral atom can be written as
\begin{equation}
n(r)=\frac{Z^{2}}{4\pi a^{3}}\left(\frac{\Phi}{x}\right)^{3/2},
\label{bu1}
\end{equation}
where $a=(1/2)(3\pi/4)^{2/3}$ and $x=Z^{1/3}r/a$, 
and the dimensionless TF differential equation is
\begin{equation}
\frac{d^{2}\Phi(x)}{dx^{2}}=\sqrt{\frac{\Phi^{3}(x)}{x}},\;\;\;\;\;\;\; \Phi(x)>0,
\label{e343}
\end{equation}
which satisfies the following initial conditions:
\ben
\Phi(0) = 1\,, ~~~~\Phi'(0)=-B\,, ~~~~ B=1.5880710226\,.
\label{eq:initcond}
\een
We construct a model for $\Phi$ which recovers
the first eight terms of the small-$x$ expansion and the leading
term of the asymptotic expansion at large-$x$
($\Phi(x)$ $\rightarrow$ $144/x^{3}$, as $x\rightarrow\infty$). Following Tal and Levy
\cite{TL81}, we use $y=\sqrt{x}$ as the variable, because of the singularity of the TF 
equation. Our parametrization is
\vspace{0.6cm}
\begin{figure}[htbp]
\begin{center}
\includegraphics[height=5.5cm]{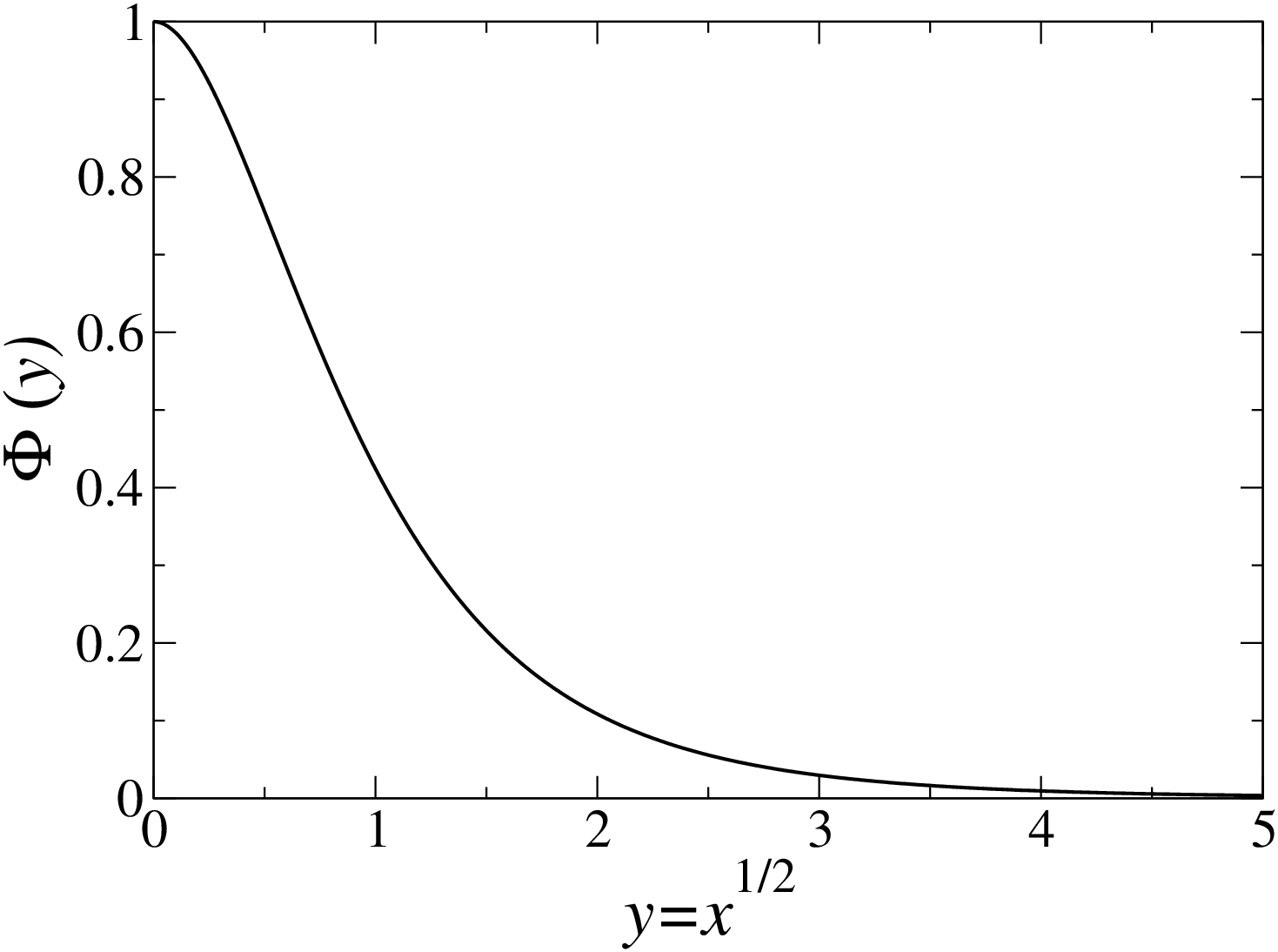}
\caption{The exact numerical $\Phi(y)$ and parametrized $\Phi(y)$ can not be distinguished.}
\label{f:PhiExact}
\end{center}
\end{figure}
\begin{figure}[htbp]
\begin{center}
\includegraphics[height=5.5cm]{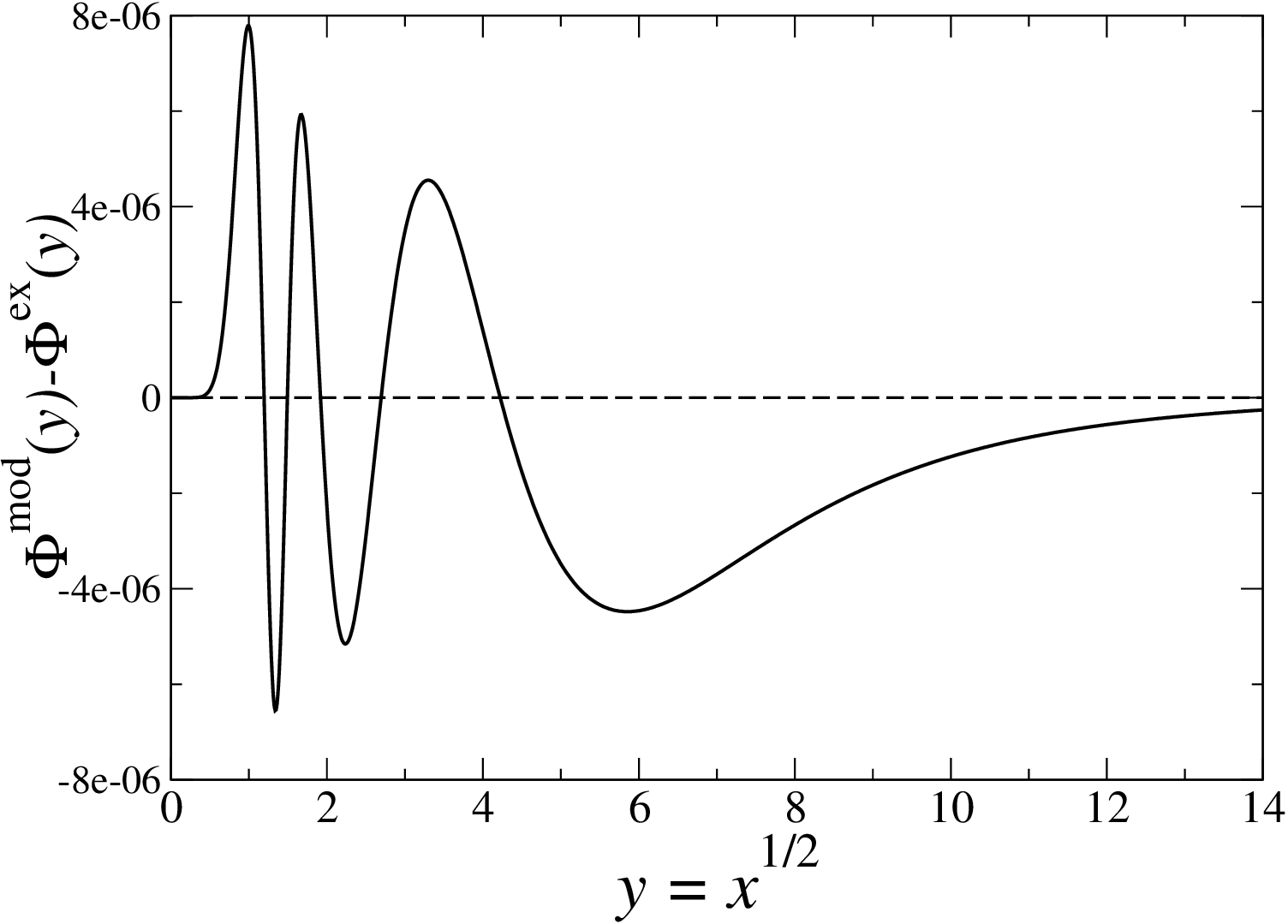}
\caption{Errors in the model, relative to numerical integrations.}
\label{f:diffphi}
\end{center}
\end{figure}
\ben
\Phi^{mod}(y) = \left(1+\sum_{p=2}^{9} \alpha_{p}y^{p}\right)/\left(1+y^{9}\sum_{p=1}^{5} \beta_{p}y^{p} + \frac{\alpha_{9}y^{15}}{144}\right)
\label{bu3}
\een
where $\alpha_{i}$ and $\beta_{i}$ are coefficients given in the Table \ref{t:Phicoeff}.
The values of $\alpha_{i}$ are fixed by the small $y$-expansion, while those of $\beta_{i}$ 
are found by minimization
of the weighted sum of squared residuals, $\chi^{2}$, for $0<y<10$. 
The $\chi^{2}$ was minimized using the 
Levenberg-Marquardt method \cite{PFTV92}. 
This method is for fitting when the model depends nonlinearly on the 
set of unknown parameters. 
1000 points were used, equally spaced between $y=0$ and $y=10$.
\begin{table}[htb]
\begin{tabular}{|c|c|c|c|}
\hline
$\alpha_{2}$ & $-B$ & $\beta_{1}$ & $-0.0144050081$ \\ 
$\alpha_{3}$ & $4/3$ & $\beta_{2}$ & $0.0231427314$ \\ 
$\alpha_{5}$ &$-2B/5$ & $\beta_{3}$ & $-0.00617782965$ \\ 
$\alpha_{6}$ & $1/3$ & $\beta_{4}$ & $0.0103191718$ \\ 
$\alpha_{7}$ & $3B^{2}/70 $ & $\beta_{5}$ & $-0.000154797772$ \\ 
$\alpha_{8}$ & $-2B/15$ &  &  \\ 
$\alpha_{9}$ & $2/27 + B^{3}/252 $ &  & \\ 
\hline
\end{tabular}
\caption{\label{t:Phicoeff} The values of  $\beta_{i}$ are found by 
fitting Eq.(\ref{bu3}) to the exact solution, and
those of  $\alpha_{i}$ are the parameters of small-$y$ 
expansion \cite{TL81}. $B$ is given by 1.5880710226.}
\end{table}
We plot the numerically exact $\Phi(y)$ and our model in Figure \ref{f:PhiExact}, 
and the differences between them in Figure \ref{f:diffphi}. These graphs illustrate 
the accuracy of our parametrization.

In Table \ref{table2} we calculate several moments using our model and existing models 
that were proposed by Gross and Dreizler \cite{GD79} and Latter \cite{LB55}. 
The Latter parametrization is
\bea
\Phi^{L}(x)&=&1/(1+0.02747x^{1/2}+1.243x-0.1486x^{3/2}\nonumber \\
&&+0.2303x^{2}+0.007298x^{5/2} \nonumber \\
&&+0.006944x^{3})\,,
\label{eqn:latter}
\eea
and the Gross-Dreizler model (which correctly removes the $\sqrt{x}$ term) is:
\bea
\Phi^{GD}(x)&=&1/(1+1.4712x-0.4973x^{3/2}+ \nonumber \\
&&0.3875x^{2}+0.002102x^{3})\,.
\label{eqn:GDmodel}
\eea
Lastly, we introduce an extremely simple model that we have found 
useful for pedagogical purposes (even when $N$ differs from $Z$).
We write 
\ben
n^{ped}(r) = \frac{N}{2\pi^{3/2}R^{3/2}}\frac{1}{r^{3/2}}\, e^{-r/R}\,, ~R= \frac{\alpha N^{2/3}}{Z-\beta N}\,,
\label{eq:vTFn}
\een
where $\alpha=(9/5\sqrt{5})(\sqrt{3}\pi/4)^{1/3}$ and $\beta=1/2-1/\pi$ have been
found from integration of the TF kinetic and Hartree
energies, respectively, and $R$ minimizes the TF total energy.
For $N=Z$, this yields:
\ben
\Phi^{ped}(x) = \gamma\, e^{-2a(1-\beta)x/3\alpha}\,,~~\gamma=\frac{5\sqrt{5}}{6\sqrt{3}}\left(\frac{1}{2}+\frac{1}{\pi}\right) \,.
\label{eq:vTFphi}
\een
This crude approximation does not satisfy the correct initial conditions of 
Eq. (\ref{eq:initcond}):
\bea
\Phi^{ped}(0)&=&\gamma=0.880361\, (\neq 1)\,, \nonumber \\
\Phi^{ped}{'}(0)&=&-\frac{125(2+\pi)^{2}}{648(4\pi^{5})^{1/3}}=-0.48~(\neq -1.59)\,.
\label{eq:vTFphi1}
\eea

To compare the quality of the various parametrizations, we 
calculate the $p$-th moment of the $j$-th power of $\Phi(x)/x$:
\ben
M^{(p)}_{j} = \int dx\, x^{p} \left( \frac{\Phi(x)}{x}\right)^{j}\, .
\label{eq:mpjdef}
\een
Many quantities of interest can be expressed in terms of these moments:

1) Particle number: To ensure $\int d^{3}r\, n(\br) = N$, we require

\begin{equation}
M^{(2)}_{3/2}=1
\label{bu14}
\end{equation}

2) TF kinetic energy: The TF kinetic energy is $c_{0}Z^{7/3}$, which implies

\begin{equation}
M^{(2)}_{5/2}=\frac{5}{7}B\,.
\label{bu15}
\end{equation} 

3) The Hartree energy is $U=\frac{1}{2}\int\int d^{3}r\,d^{3}r'\, \frac{n(\br)n(\br')}{|\br-\br'|}=\frac{1}{7a}M^{(1)}_{3/2}Z^{7/3}$, which implies 

\ben
M^{(1)}_{3/2}=B\,.
\label{bu16}
\een

4) The external energy is defined as $V\ext=-\int d^{3}r\, Z\,n(r)/r=-\frac{1}{a}M^{(1)}_{3/2}Z^{7/3}$ 
for the exact TF density, which also implies Eq. (\ref{bu16}).

5) The local density approximation (LDA) exchange 
energy is defined as $E\x\LDA=A\x\int d\mathbf{r}\, n^{4/3}(\br)$,
where $A\x = -(3/4)(3/\pi)^{1/3}$,
so for TF, $E\x\LDA=A\x(4\pi a^{3})^{(-1/3)}M^{(2)}_{2}Z^{5/3}$, which implies

\begin{equation}
M^{(2)}_{2} = 0.615434679\,.
\end{equation}
This $M^{(2)}_{2}$ is evaluated on the exact TF density which we
calculate numerically. LDA exchange suffices \cite{Sa80,PCSB06} for asymptotic
exactness to the order displayed in Eqs. (\ref{EZasymp}) and (\ref{TsZasymp}); 
for a numerical study, see Ref. \cite{KLSS97}.
\begin{table*}[htbp]
\caption{ Various moments calculated with our model and with the models of Ref. \cite{GD79, LB55}.
Here $M^{(p)}_{j}$ is given by $\int dx\, x^{p} \left ( \frac{\Phi (x)}{x} \right )^{j}$  } 
\begin{tabular}{|c|c|c|c|c|c|c|c|c|c|}
\hline
moment & our model & \% error & Gross and Dreizler \cite{GD79}& \% error & Latter\cite{LB55} & \% error & $\Phi^{ped}(x)$ & \% error & exact\\ \hline
$M^{(2)}_{3/2}$ & 0.999857885& -0.01 & 1.008 & 0.8 & 0.999 & -0.04 & 1 & 0 &1 \\ \hline
$M^{(2)}_{5/2}$ & 1.13426462 & -0.006 & 1.1299 & -0.4&  1.137 & 0.2 & 1.11 & -2 &$5B/7$ \\ \hline
$M^{(2)}_{2}$ & 0.615438208 & 0.001 & 0.6129 & -0.4 & 0.616 & 0.02 & 0.72 & 16 &0.615434679\footnotemark[1] \\ \hline
$M^{(1)}_{3/2}$ & 1.58799857&  -0.005 & 1.5844 & -0.2 & 1.589 & 0.07 & 1.62 & 2 &B \\ \hline
\hline
\end{tabular}
\footnotetext[1]{Numerical result from the TF differential equation.}
\label{table2}
\end{table*}
Table \ref{table2} shows that our modern parametrization is far more accurate
than existing models by all measures, and that our simple pedagogical model is
roughly correct for many features.

Finally, we make some comparisons with densities of real atoms
to illustrate those features of real atoms that
are captured by TF.
The radial density, $s(\br)$ (Eq. (\ref{s})), and $q(\br)$ (Eq. (\ref{q2})) are given by
\ben
4\pi r^{2}n(r) = Z^{4/3}f(x)/a\,,
\label{eq:TFdens}
\een
where $f(x)=\sqrt{x}\, \Phi^{3/2}(x)$,
\ben
s(r) = \frac{a_{1}}{Z^{1/3}} \frac{|g(x)|}{f(x)} \,,~ a_{1}= ( 9/2\pi) ^{1/3} /2 \, ,
\label{eq:TFs}
\een
and
\ben
q(r) = \frac{a^{2}_{1}}{3Z^{2/3}}\frac{\{g^{2}(x)+2x^{2}\Phi(x)\Phi''(x)\}} 
{f^{2}(x)} \,,
\label{eq:TFq}
\een
where $g(x)$ is defined as $\Phi(x)-x\Phi'(x)$.
The gradient relative
to the screening length is
\ben
t(\br) = \frac{|\nabla n(\br)|}{2k\s (\br)n(\br)}\,,~\text{where}~ k\s(\br) = \sqrt{4k\F (\br) / \pi}\,, 
\een
and here
\ben
t(r)= \frac{a_{2}|g(x)|}{(x^{3}\Phi^{5}(x))^{1/4}} \,,~a_{2} =  \frac{3^{5/6}\pi^{1/3}}{2^{8/3}\sqrt{a}} = 0.6124\,.
\label{eq:TFt}
\een

We also show large- and small-$x$ limit behaviors of various quantities
using $\Phi(x) \to 144/x^{3}$ as $x \to \infty$ and $\Phi(x) \to 1-Bx+\cdots$ as $x \to 0$.
\bea
\frac{Z^{2}}{4\pi a^{3}}\frac{1}{x^{3/2}} ~\stackrel{x \to 0}{\longleftarrow} &n(r)& \stackrel{x \to \infty}{\longrightarrow}~ \frac{432Z^{2}}{a^{3}\pi x^{6}}\,, \\
\frac{Z^{4/3}}{a} \sqrt{x} ~\stackrel{x \to 0}{\longleftarrow}  &4\pi r^{2}n(r)& \stackrel{x \to \infty}{\longrightarrow}~ \frac{144Z^{4/3}}{ax^{5/2}}\,, \\
\frac{a_{1}}{Z^{1/3}} \frac{1}{\sqrt{x}} ~\stackrel{x \to 0}{\longleftarrow} &s(r)& \stackrel{x \to \infty}{\longrightarrow}~ \frac{a_{1}x}{3Z^{1/3}}\,, \\
\frac{a^{2}_{1}}{3Z^{2/3}} \frac{1}{x} ~\stackrel{x \to 0}{\longleftarrow} &q(r)& \stackrel{x \to \infty}{\longrightarrow}~ \frac{5a^{2}_{1}x^{2}}{54Z^{2/3}}\,, \\
\frac{a_{2}}{x^{3/4}} ~\stackrel{x \to 0}{\longleftarrow} &t(r)& \stackrel{x \to \infty}{\longrightarrow} ~ \frac{2a_{2}}{\sqrt{3}}\,.
\eea
\vspace{0.6cm}
\begin{figure}[htbp]
\begin{center}
\includegraphics[height=5.5cm]{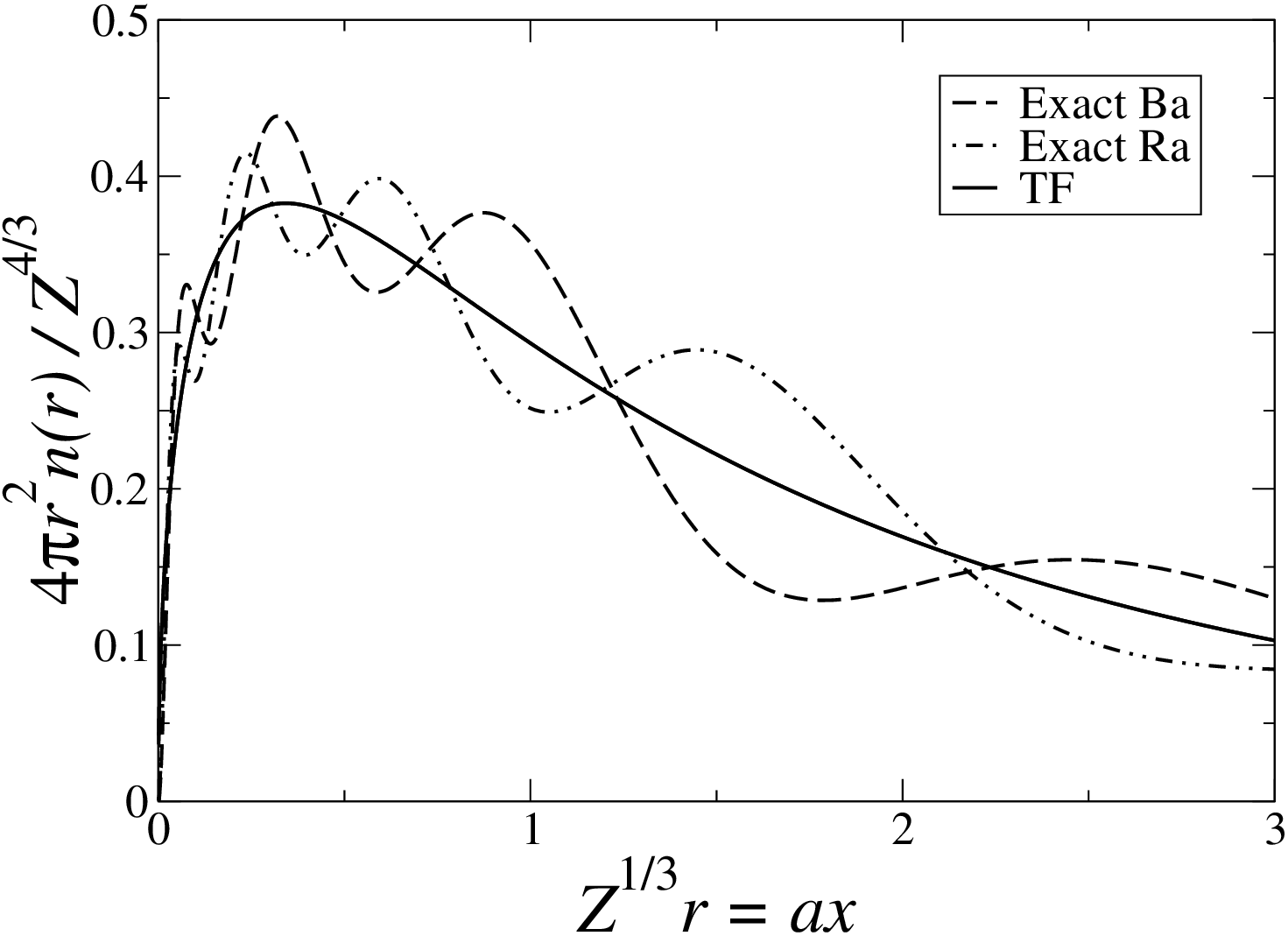}
\caption{Plot of the scaled radial densities of Ba and Ra using Eq.(\ref{eq:TFdens}) and SCF densities with OEP exact exchange.
TF scaled densities of Ba and Ra are on top of each other.}
\label{f:rd}
\end{center}
\end{figure}
\begin{figure}[htbp]
\begin{center}
\includegraphics[height=5.5cm]{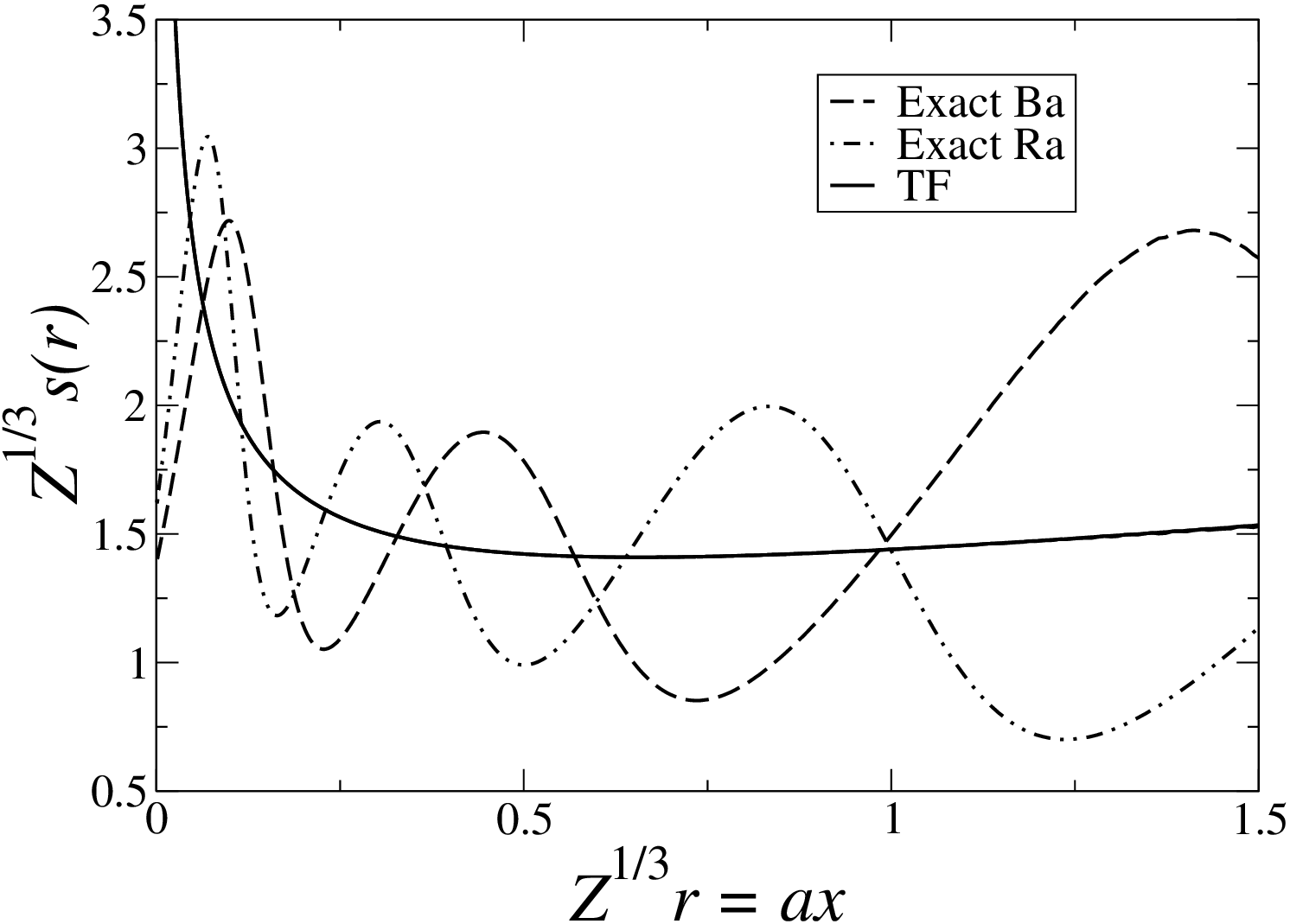}
\caption{Plot of the scaled reduced density gradient $s(r)$ (relative to the local
Fermi wavelength) vs. $Z^{1/3} r$.}
\label{f:s2}
\end{center}
\end{figure}

We plot the $Z$-scaled exact (self-consistent densities with OEP exact exchange functional)
and TF radial densities of Ba ($Z=56$) and Ra ($Z=88$) in Figure \ref{f:rd}. Although the shell structure is missing, 
and the decay at a large distance is wrong, the overall shape of the TF density is relatively correct.

\begin{figure}[htbp]
\begin{center}
\includegraphics[height=5.5cm]{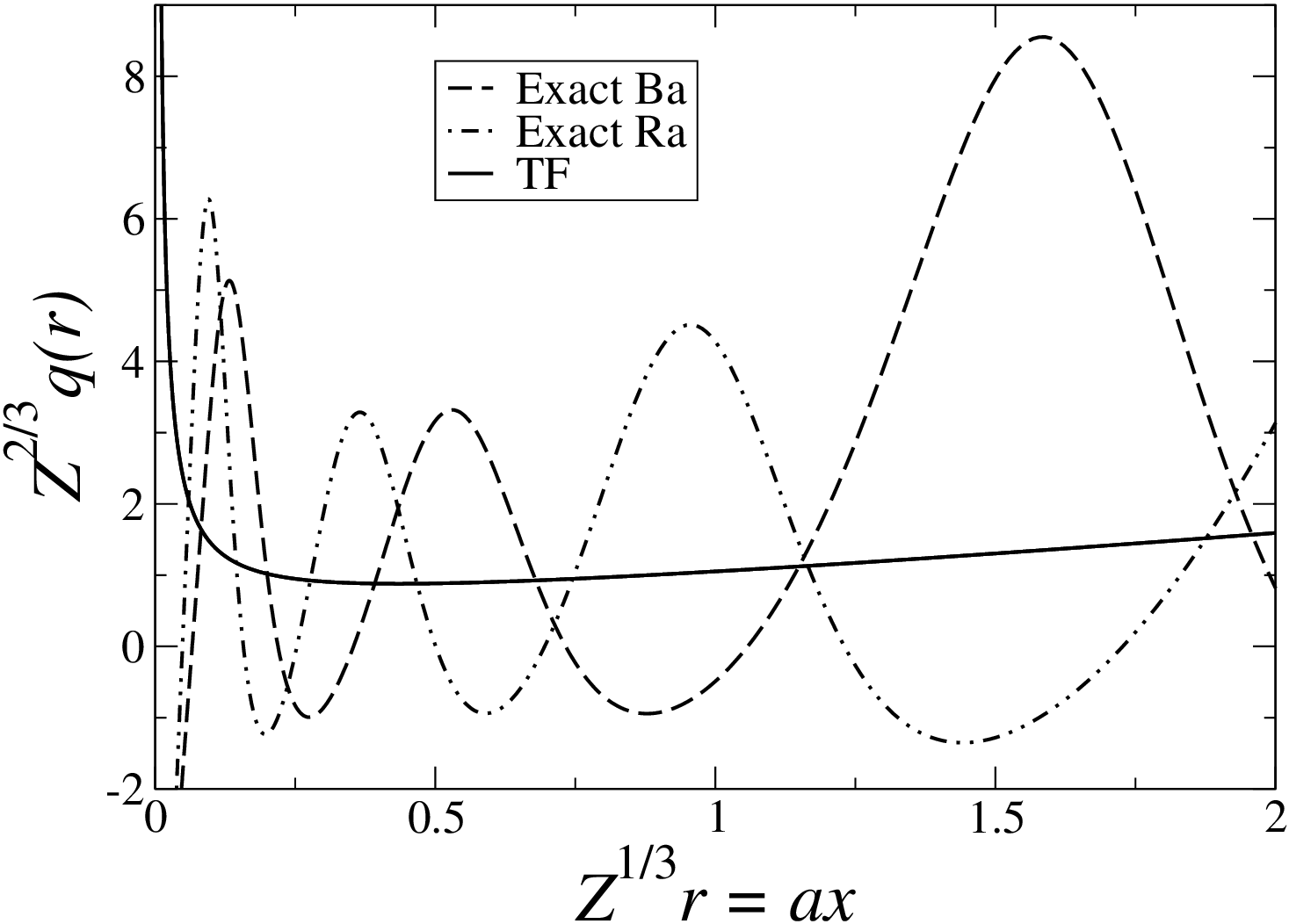}
\caption{Plot of the scaled reduced Laplacian $q(r)$ (relative to the local Fermi wavelength)
vs. $Z^{1/3} r$.}
\label{f:q2}
\end{center}
\end{figure}

\begin{figure}[htbp]
\begin{center}
\includegraphics[height=5.5cm]{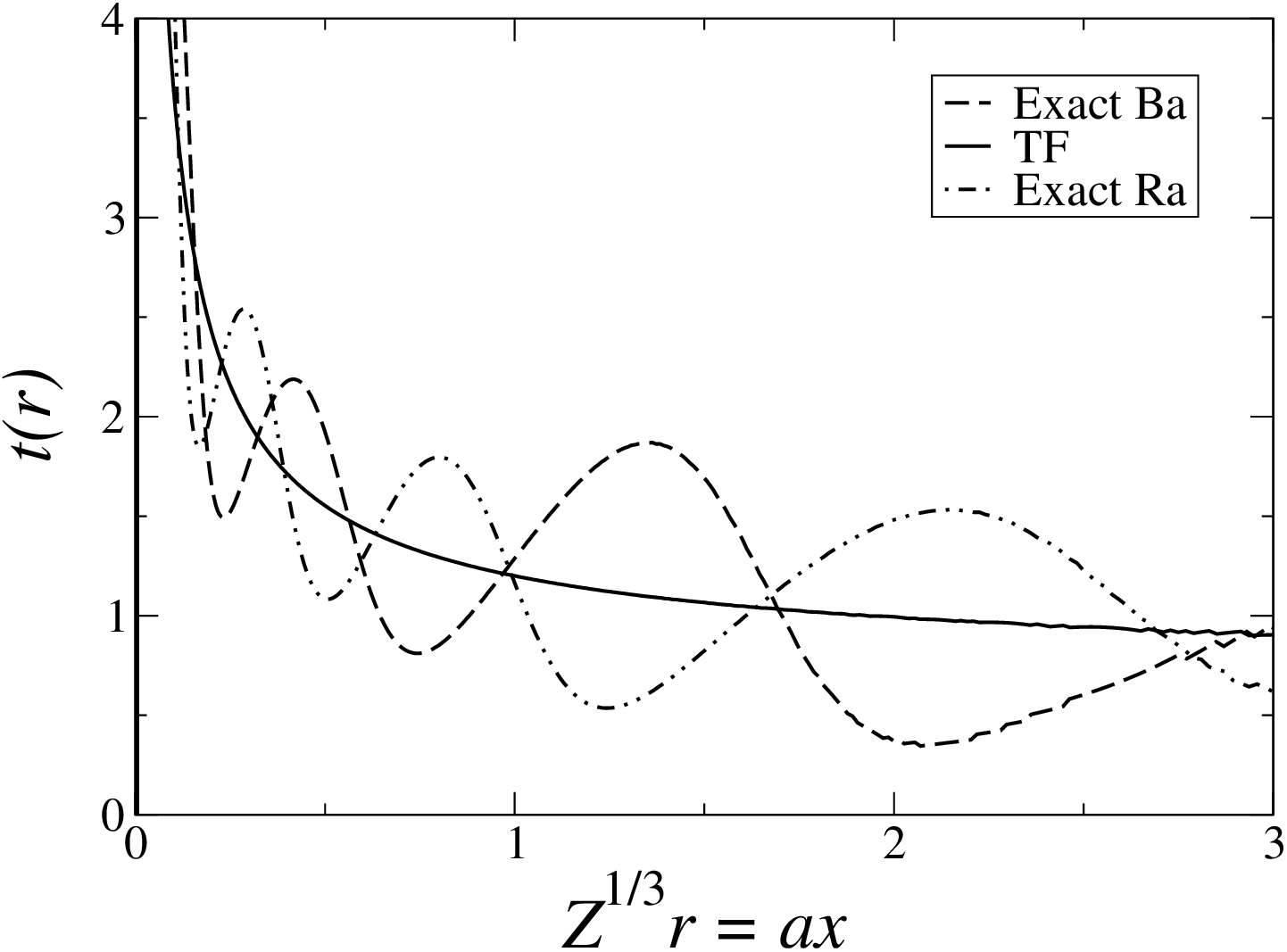}
\caption{Plot of the reduced density gradient $t(r)$ (relative to the local screening length) vs. $Z^{1/3} r$. As $r \to \infty$, the TF $t \to 0.7071$.}
\label{f:t2}
\end{center}
\end{figure}

In Figures \ref{f:s2}, \ref{f:q2}, and \ref{f:t2},
we plot the scaled $s(r)$, $q(r)$, and $t(r)$ using the exact
and the TF densities of Ba and Ra.
In particular, $t(r)$ measures how fast the density changes on the scale of
the TF screening length, and its magnitude does not 
vary with $Z$ in TF theory.
From these figures, we see that $s(r)$, $q(r)$ and $t(r)$ of the TF density 
diverge near the nucleus, since the TF density does not satisfy Kato's cusp condition. 
\vspace{0.6cm}

When $N=Z \to \infty$ for a realistic density, $s(r)$ is small except in the density tail 
($s \sim Z^{-1/3}$ over most of the density), 
and $q(r)$ is small except in the tail and $1s$ core regions ($q \sim Z^{-2/3}$ 
over most of the density). 
This is why gradient expansions for the kinetic and exchange energies, 
applied to realistic densities, work as well as they do in this limit. 
The kinetic and exchange energies have only one 
characteristic length scale, the local Fermi wavelength, 
but the correlation energy also has a different one,
the local screening length.  Since $t(r)$ is not and does not become 
small in this limit,
gradient expansions do not work well at all 
for the correlation energies of atoms \cite{PCSB06}.
The standard of ``smallness" for $s$ and $q$, and the more severe standard
of smallness for $t$, are explained in
Refs. \cite{PCSB06} and \cite{PRCV08}.

Finally we evaluate $T^{(0)}+T^{(2)}$ on the TF density. We find the 
correct $c_0$ in the $Z \rightarrow \infty$ expansion from $T^{(0)}$, but
$c_1$ vanishes, due to the absence of a proper nuclear cusp, 
and $c_2$ diverges because $T^{(2)}$
diverges at its lower limit of integration.

\section{Summary}
\label{s:sum}
We have shown the importance of the large-$N$ limit for
density functional construction of the kinetic energy (with the functional evaluated
on a Kohn-Sham density), and also provided a modern,
highly accurate parameterization of the neutral-atom TF 
density. Our results should prove useful in the never-ending search
for improved density functionals.
\vspace{0.9cm}

For atoms and molecules, the large-$N$ limit seems more important than the slowly-varying limit.
On the ladder \cite{PS01} of density-functional approximations, 
there are three rungs of semilocal approximations 
(followed by higher rungs of fully nonlocal ones).  
The LDA uses only the local density, 
the GGA uses also the density
gradient, and the meta-GGA uses in addition the orbital kinetic 
energy density or the Laplacian of the density.  
For the exchange-correlation energy, the GGA rung 
cannot \cite{PCSB06, PRCV08} simultaneously describe the slowly-varying 
limit and the $N=Z \to \infty$ limit for an atom, and we have found 
here that the same is true (but less severely by percent error 
of a given energy component) for the kinetic energy. 
This follows because, as $N=Z \to \infty$, the reduced 
gradient $s(r)$ of Eq. (\ref{s}) becomes small over the energetically 
important regions of the atom, as can be inferred from Fig. \ref{f:s2}, 
so that a GGA reduces to its own second-order gradient
expansion even in regions where a meta-GGA does not \cite{PCSB06}
(e.g., near a nucleus,
where $q(r)$ diverges but $s(r)$ does not, as shown in Figs. \ref{f:s2} and \ref{f:q2}).
For the kinetic as for
the exchange-correlation energy, meta-GGA's \cite{PC07} can
recover both the slowly-varying and large-$Z$ limits;
it remains to be seen how well fully nonlocal approximations \cite{WC00, GA08}
can do this.

We thank Eberhard Engel for the use of his atomic OPMKS code, and NSF 
(CHE-0355405 and DMR-0501588) and the Korea Science and Engineering 
Foundation Grant (No. C00063), for support.

\end{document}